\documentclass[final]{svjour2}
\usepackage{graphicx}
\usepackage{rotating}
\usepackage{amssymb}
\usepackage{mathptmx}
\usepackage{color}
\usepackage[numbers]{natbib}
\makeatletter
\journalname{Journal of Low Temperature Physics}
\bibpunct{}{}{,}{s}{}{,}

\begin{document}

\newcommand{\hdblarrow}{H\makebox[0.9ex][l]{$\downdownarrows$}-}
\newcommand{\WHz}{\mathrm{W\ Hz^{-0.5}}}
\newcommand{\mum}{\mathrm{\mu m}}
\newcommand{\ee}[1]{\times 10^{#1}}

\title{Power Handling and Responsivity of Submicron Wide Superconducting Coplanar Waveguide Resonators}

\author{R.M.J.~Janssen$^1$ \and A.~Endo$^1$ \and J.J.A.~Baselmans$^2$ \and P.J.~de~Visser$^{1,2}$ \and R.~Barends$^3$ \and T.M.~Klapwijk$^1$}

\institute{1:Physics of NanoElectronics Group, Kavli Institute of Nanoscience,\\ Faculty of Applied Sciences, Delft University of Technology \\ Lorentzweg 1, 2628CJ Delft, The Netherlands. \email{r.m.j.janssen@tudelft.nl}
\\2:SRON Netherlands Institute for Space Research,\\ Sorbonnelaan 2, 3584CA Utrecht, The Netherlands
\\3:Department of Physics, University of California, Santa Barbara, CA 93106, USA}

\date{Recieved: 17 July 2011 / Accepted: 2 January 2012 / Published online: 20 January 2012}

\maketitle


\begin{abstract}

The sensitivity of microwave kinetic inductance detectors (MKIDs) based on coplanar waveguides (CPWs) needs to be improved by at least an order of magnitude to satisfy the requirements for space-based terahertz astronomy. Our aim is to investigate if this can be achieved by reducing the width of the CPW to much below what has typically been made using optical lithography ($>1$ $\mum$). CPW resonators with a central line width as narrow as 300 nm were made in NbTiN using electron beam lithography and reactive ion etching. In a systematic study of quarter-wave CPW resonators with varying widths it is shown that the behavior of responsivity, noise and power handling as a function of width continues down to 300 nm. This encourages the development of narrow KIDs using Al in order to improve their sensitivity.

\end{abstract}

\section{\label{Introduction}Introduction}
Microwave Kinetic Inductance Detectors (MKIDs)\cite{Day2003} are showing promising results to become the future of large detector arrays for terahertz astronomy. The main advantages of MKIDs are ease of fabrication, wide dynamic range and above all their inherent capability to read many pixels using frequency domain multiplexing\cite{Yates2009}. This has allowed a rapid development of MKID arrays over the past decade, which has recently resulted in the first observations using cameras based on MKID technology at ground-based (sub-)mm telescopes\cite{Schlaerth2008,Monfardini2010}. A common MKID pixel design used in these cameras is an antenna-coupled coplanar waveguide (CPW) resonator patterned in a superconducting film\cite{Day2006,Yates2009a}. The Al resonators of such MKID pixels have shown\cite{deVisser2011} a detector Noise Equivalent Power (NEP) as low as $3\ee{-19}$ $\WHz$. While this is close to the sensitivity required for background photon noise limited photometry in space, $NEP \sim 10^{-19}$ $\WHz$, an improvement of two orders of magnitude is required to achieve this for spectroscopy\cite{Benford2004}: $NEP \sim 3\ee{-21}$ $\WHz$.\\
One possible route to improve the NEP is to reduce the width of the CPW. Experiments on Al resonators wider than a few micrometers indicate that the NEP\cite{Mazin} scales as $NEP_{\theta}\propto S^{0.4}$ and $NEP_{R}\propto S^{0.7}$ for phase and amplitude read-out, respectively. This assumes $S/W$ is kept constant. Here $S$ is the CPW central line width and $W$ is the width of the CPW slots. The sensitivity improvement results from the fact that the responsivity increases more rapidly $\left( \delta x / \delta N_{qp} \propto S^{-1.7} \right)$ than the noise $\left(\sqrt{S_{\theta}}\propto S^{-1.3},\sqrt{S_{R}} \propto S^{-1} \right)$ for decreasing width. The increase in responsivity is due to a change in volume $(V \propto S)$ and kinetic inductance\cite{Porch2005,Gao2006} $(\alpha \propto S^{-0.7})$, while the increase in noise is a result of decreasing read-out power handling\cite{deVisser2010} $(P_{read} \propto S^{2})$, which also affects the Two Level System (TLS) noise\cite{Gao2007,Gao2008a} in phase read-out $(S_{\theta} \propto S^{-1.6}P_{read}^{-0.5})$. Because of the low background loading in space, the majority of the quasi-particles are expected to be the excess quasi-particles created by microwave read-out power\cite{deVisser2011} rather than those created by optical pair breaking. Therefore, the quasi-particle lifetime is unaffected by the reduced width. If these trends continue down to a width of e.g., $S=300$ nm, the NEP would reach $6\ee{-20}$ $\WHz$, which is sufficient for space-borne photometry. However, effects negligible for wide resonators could begin to play a significant role when the width approaches the film thickness and characteristic length scales such as the magnetic penetration depth.\\
In this study, we systematically investigate the behavior of noise, responsivity and read-out power handling of the CPW resonators that are as narrow as 300 nm, in order to find out if width reduction is a viable route to improve the sensitivity of MKIDs to a level suitable for space-borne applications.
\section{Experimental details}
\begin{figure}
		\hspace{0.02\textwidth}
		\begin{minipage}{0.40\textwidth}
    \centering
		\includegraphics[width=1\textwidth]{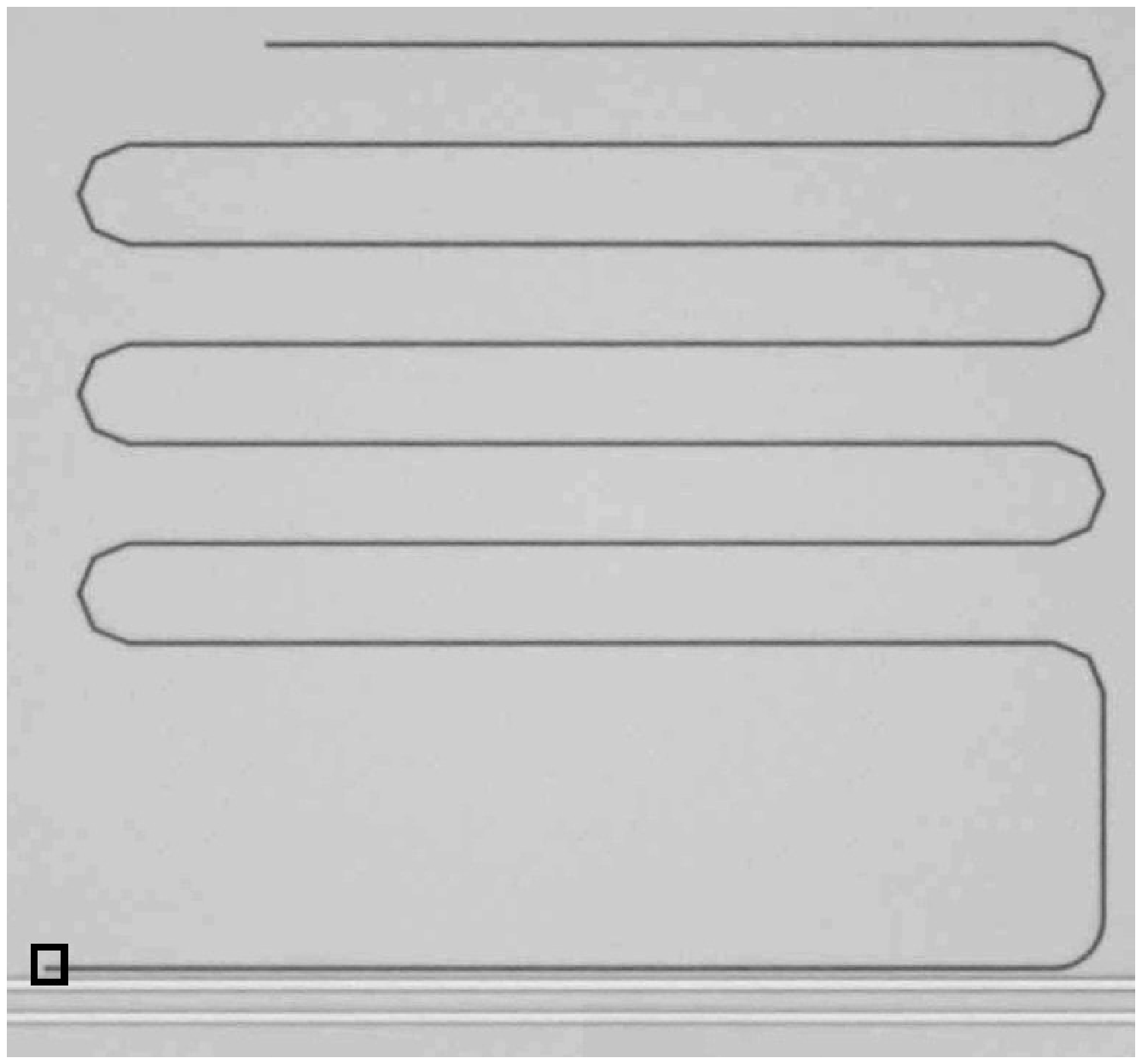}
    \end{minipage}
    \hspace{0.04\textwidth}
    \begin{minipage}{0.52\textwidth}
    \centering
    \includegraphics[width=1\textwidth]{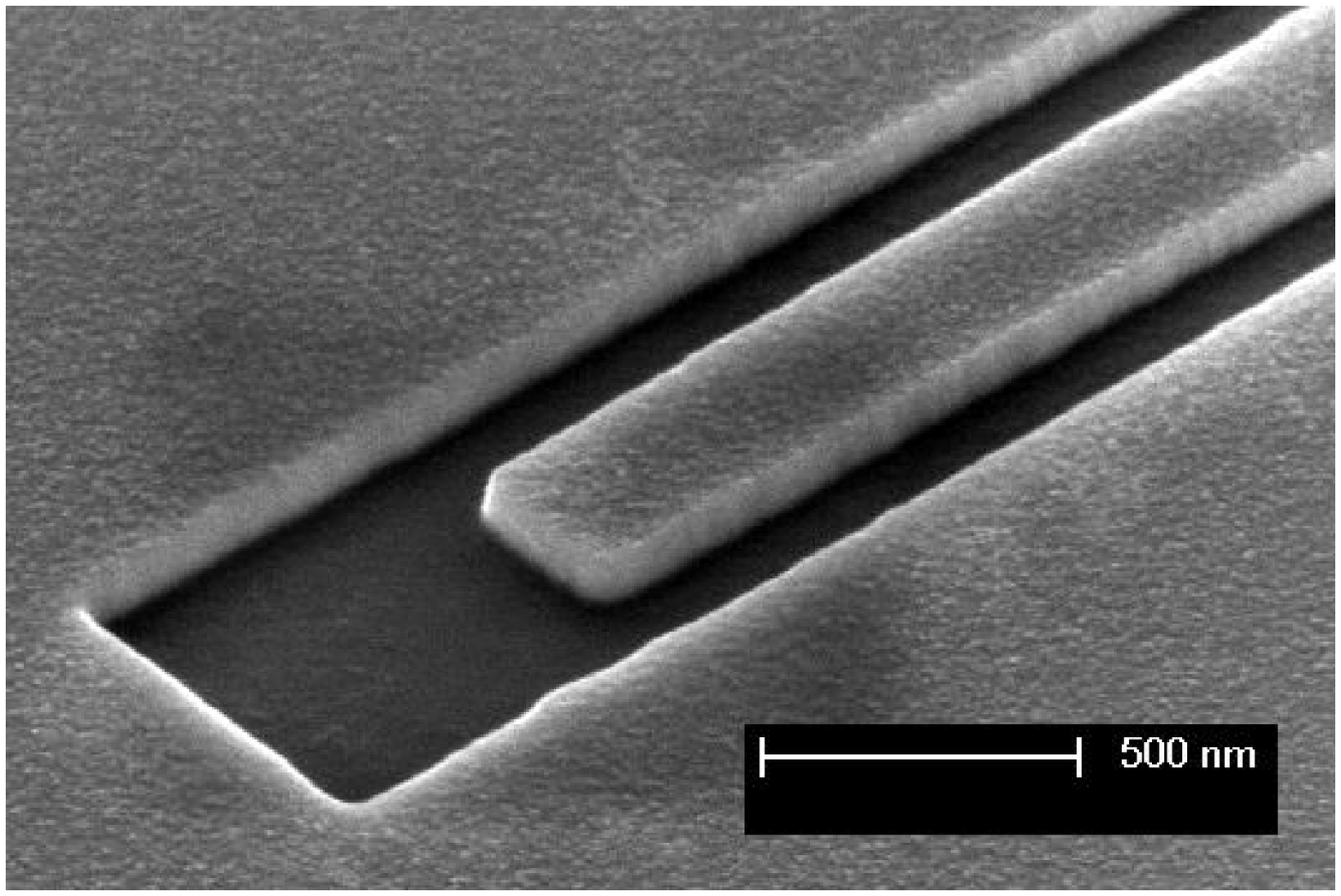}
    \end{minipage}
    \caption{\textbf{[left]} Optical micrograph of a submicron wide resonator coupled to a 22 $\mum$ wide feedline. \textbf{[right]} Zoom in on the black square of the optical image using a SEM. This SEM image shows the open end of a narrow CPW quarter-wavelength resonator patterned in 100 nm thick NbTiN on a Si substrate. The central line and slots are 300 nm and 200 nm wide, respectively.}
    \label{FigSEM}
\end{figure}

Electron beam lithography (EBL) and reactive ion etching (RIE) are used to fabricate CPW resonators with a central line width as narrow as 300 nm. Figure \ref{FigSEM} shows on the left an optical micrograph of such a narrow quarter-wavelength (QW) resonator coupled to a 22 $\mum$ wide feedline. A close-up of the open end of the resonator, outlined by the black box, can be seen in the right image of Fig. \ref{FigSEM}. The EBL system can pattern 150 of these narrow resonators in 1 hour.\\
Using only EBL and RIE, CPW resonators with a central line width, $S$, varying between 0.3 $\mum$ and 3.0 $\mum$ were patterned in a 100 nm thick NbTiN film, which was sputtered on a hydrogen passivated high resistivity ($>1$ $\mathrm{k\Omega}$ cm) $\langle$100$\rangle$-oriented Si substrate. These resonators have a length between 3.5 and 5.0 mm. The $S/W$ ratio was kept constant at $3/2$. The short quasi-particle lifetime $(\tau_{qp}\sim 1$ ns$)$\cite{BarendsThesis} of stoichiometric NbTiN prevents it from being used as the active material for MKIDs and therefore any optical experiments. However, in this work NbTiN is used instead of Al, because the higher $T_c = 13.7$ K (measured) allows measurements in a He-3 sorption cooler with a base temperature of $T_0=310$ mK. The sample is placed in this cryostat in a gold-plated Cu box that is surrounded by a superconducting shield. The feedline transmission is measured using a signal generator, low noise amplifier (LNA) and quadrature mixer\cite{Day2003,BarendsThesis,Barends2008}.\\
Detailed information on the specific design parameters and basic measurement properties of each resonator can be found in Janssen \cite{JanssenMSc}.
\section{\label{Results}Results}
\begin{figure}
\begin{center}
\includegraphics[%
  width=0.50\linewidth,
  keepaspectratio]{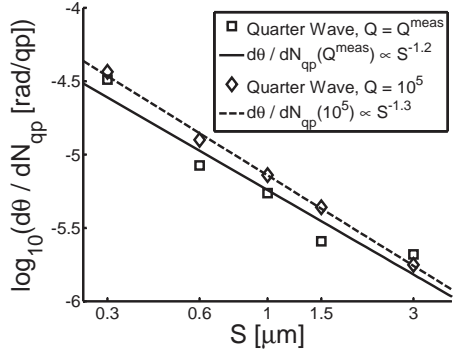}
\end{center}
\caption{The phase responsivity as a function of central line width (squares). A fit of this data shows a width dependency of $\delta \theta / \delta N_{qp} \propto S^{-1.2}$ (solid line). The main source of scatter in this data is the variation of resonator $Q_L$. For a fixed $Q_L=10^5$ (diamonds) the scatter is reduced and a fit shows a width dependency $\delta \theta / \delta N_{qp} \propto S^{-1.3}$ (dashed line).}
\label{FigResponsivity}
\end{figure}

\begin{figure}
		\begin{minipage}{0.48\textwidth}
    \centering
		\includegraphics[width=1\textwidth]{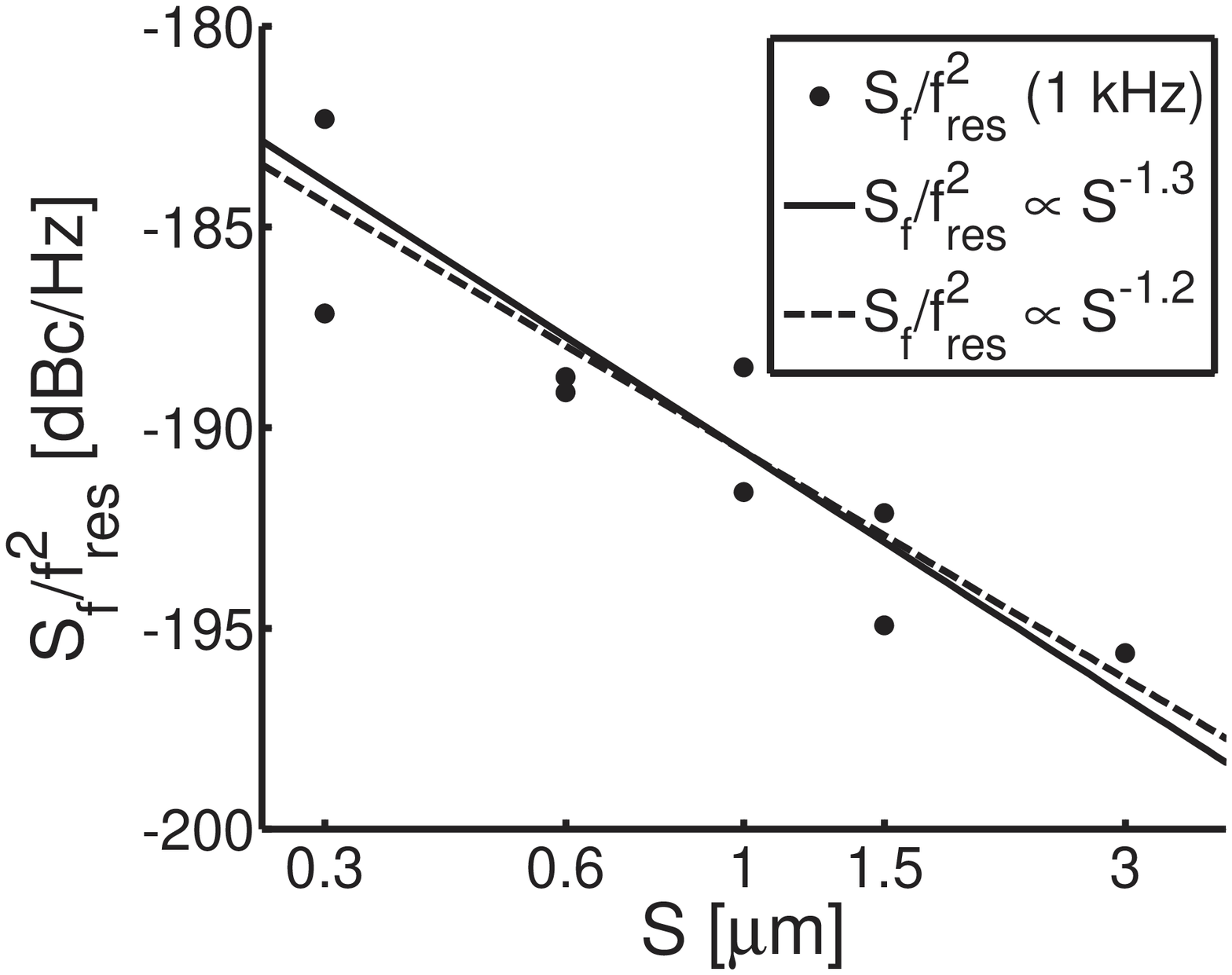}
    \end{minipage}
    \hspace{0.04\textwidth}
    \begin{minipage}{0.48\textwidth}
    \centering
    \includegraphics[width=1\textwidth]{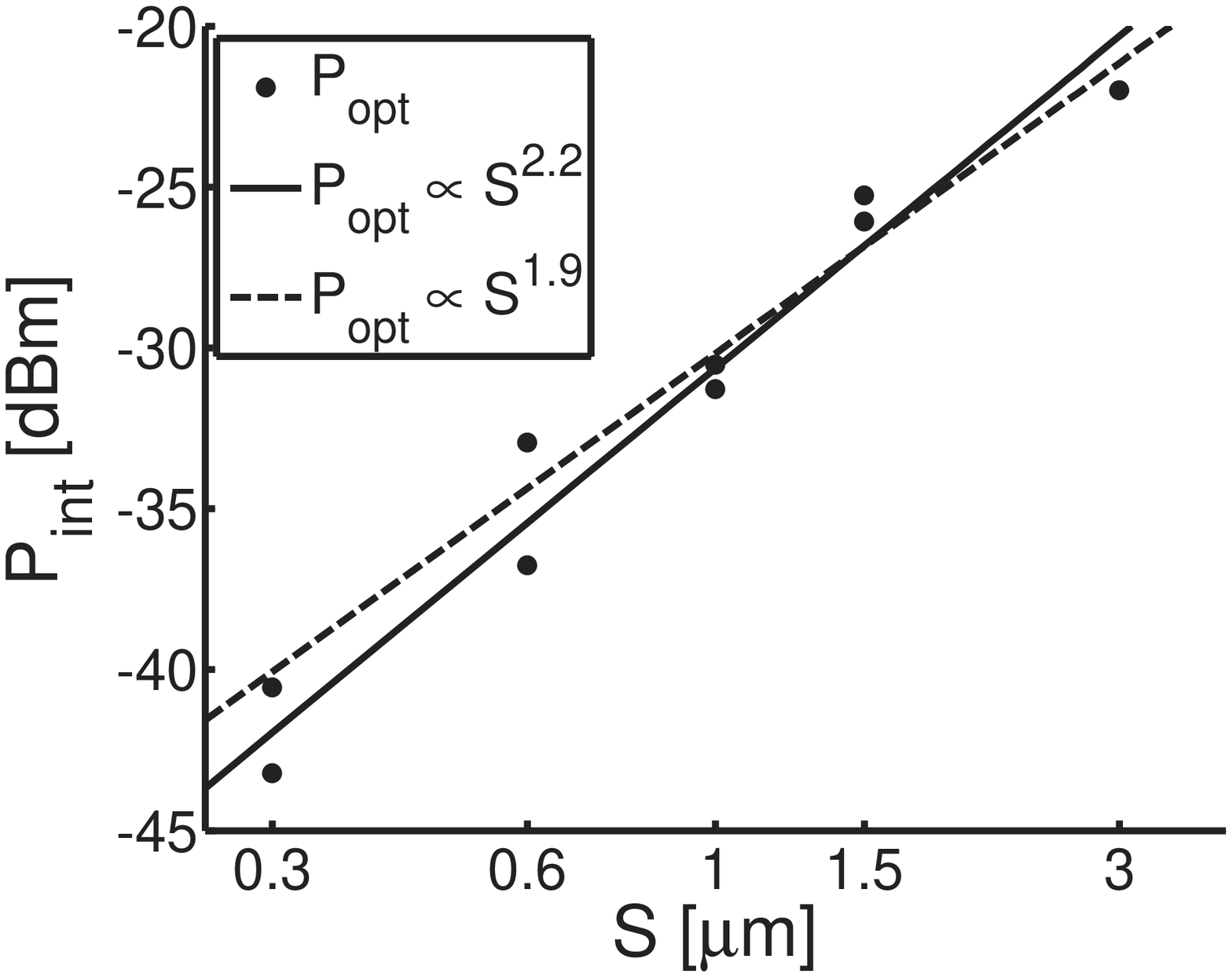}
    \end{minipage}
    \caption{\textbf{[left]} The frequency noise at 1 kHz measured at an internal power of -30 dBm as a function of central line width. A fit to the data for $S\geq1$ $\mum$ shows a width dependence $S_f/f_{res}^2 \propto S^{-1.19 \pm 0.50}$ (dashed line), while a fit to all measurements results in $S_f/f_{res}^2 \propto S^{-1.28 \pm 0.21}$ (solid line). \textbf{[right]} The internal power at the measured optimum read-out power as a function of central line width. A fit to the data for $S\geq1$ $\mum$ shows a width dependence $P_{int} \propto S^{1.89 \pm 0.34}$ (dashed line), while a fit to all measurements results in $P_{int} \propto S^{2.16 \pm 0.20}$ (solid line)}
    \label{FigNoisePower}
\end{figure}

Following Baselmans \textit{et al.}\cite{Baselmans2008} the phase responsivity of each resonator is determined from the change in resonance frequency, $f_{res}$, as a function of temperature, $T$. 
\begin{equation}
\frac{\delta \theta}{\delta N_{qp}} = -4Q_L\frac{\delta y}{\delta N_{qp}}
\label{6_PhaseFreq}
\end{equation}
where $Q_L$ is the resonator's loaded quality factor, $y =(f_{res}(T)-f_{res}(T_0))/f_{res}(T_0)$ and $N_{qp}$ is the number of quasi-particles in the resonator volume. The last quantity is given by the temperature under the assumption of a steady-state number density of thermally excited quasi-particles\cite{Kaplan1976}. A linear least squares (LLS) fit is applied for temperatures $T > T_c/6$ to determine $\delta y/\delta N_{qp}$. Figure \ref{FigResponsivity} shows that the phase responsivity increases for decreasing widths (squares). The dependency can be described by a power law $\delta \theta / \delta N_{qp} \propto S^{-1.20\pm 0.19}$ (solid line). The scatter in Fig. \ref{FigResponsivity} is mainly caused by the variation in the measured loaded quality factor, $Q_L^{meas}$, between different resonators. This is easily shown by substituting $Q_L=10^5$, which for these resonators is a typical measured value $\langle \log_{10}Q_L^{meas} \rangle=4.90 \pm 0.12$, into Eq. \ref{6_PhaseFreq}. The responsivities calculated thus are shown by the diamonds in Fig. \ref{FigResponsivity}. Application of a constant $Q_L$ only changes the slope of the LLS fit slightly: $\delta \theta / \delta N_{qp} \propto S^{-1.29\pm0.04}$ (dashed line), but the reduced scatter improves the uncertainty of the fit significantly. Since all resonators were overcoupled a reduced scatter should be achievable by improving the coupler design to achieve a more uniform coupling quality factor, $Q_c$.\\
Besides responsivity, the sensitivity of a resonator is determined by the noise. In phase read-out the noise can be quantified by the power spectral density (PSD) of the $Q_L$-independent frequency noise,  $S_f/f_{res}^2$. The left graph in Fig. \ref{FigNoisePower} shows the value of the frequency noise PSD at 1 kHz measured at an internal power $P_{int}=-30$ dBm as a function of central line width. A LLS fits to the data for $S\geq1$ $\mum$ shows a width dependency $S_f/f_{res}^2 \propto S^{-1.19\pm0.50}$ (dashed line). All measurements for narrower resonators are within 3 dBc/Hz of this line. A LLS fit using all available data gives us a width dependency $S_f/f_{res}^2 \propto S^{-1.28\pm0.21}$ (solid line).\\
Using a higher read-out power reduces the TLS frequency noise. However, at too high a read-out power a resonator will show unwanted non-linear behavior. Hence each resonator has an optimum read-out power which balances these phenomena. We define the optimum power as the read-out power at which the transmission at resonance is not showing any non-linear behavior and the mean amplitude noise between 10 and 100 Hz is the lowest. This minimum in the noise exists because an increase in read-out power will reduce the relative noise level of the LNA, while overdriving manifests itself as additional $1/f$-noise. From the optimum read-out power, $P_{read}$, the corresponding internal power, $P_{int}$ is determined using\cite{BarendsThesis}:
\begin{equation}
P_{int} \cong \frac{2Q_L^2}{\pi Q_c}\frac{Z_{feedline}}{Z_{res}}P_{read}
\label{3_Pconversion}
\end{equation}
where $Z_{feedline}$ and $Z_{res}$ are the impedances of the feedline and resonator, respectively. The measured optimum internal power as a function of the central line width is shown in the right graph of Fig. \ref{FigNoisePower}. A LLS fit shows a power law dependence $P_{int} \propto S^{2.16 \pm 0.20}$ (solid line). When applying a LLS fit to the measurements with $S\geq1$ $\mum$ a dependency $P_{int} \propto S^{1.89 \pm 0.34}$ (dashed line) is found. Both these power laws are consistent with the maximum read-out power being limited by the current density in the resonator, which would give $P_{int} \propto S^{2}$. While the mechanism that limits the power handling of a resonator has not been identified, all likely candidates depend on the microwave current density\cite{deVisser2010}.
\section{Discussion}
As shown by the results presented in Sect. \ref{Results} the measurements of the narrow resonators $(S<1$ $\mum )$ show no deviation from the trends set by the wider resonators $(S\geq 1$ $\mum )$ for both frequency noise and optimum internal power. The scatter around the LLS fit of both quantities are typical variations caused during fabrication. While a real NEP cannot be determined for NbTiN because we lack lifetime data, a width dependence can be found using the power laws found in Sect. \ref{Results}. Based on this work a negligible improvement of the NEP is found: $NEP_{\theta} \propto S^{0.11 \pm 0.27}$ and $NEP_{R} \propto S^{0.21 \pm 0.11}$. These values are in line with the expected improvement of NbTiN resonators: $NEP_{\theta} \propto S^{0.0}$ and $NEP_{R} \propto S^{0.3}$. The improvement of NbTiN resonator is less than that expected for narrow Al resonators, because the responsivity of NbTiN resonators does not improve as rapidly. For NbTiN resonators a $\delta x/\delta N_{qp} \propto S^{-1.3}$ is expected and measured, while $\delta x/\delta N_{qp} \propto S^{-1.7}$ is expected for resonators in Al, as outlined in Sec. \ref{Introduction}. This difference exists, because the large magnetic penetration depth gives NbTiN a higher kinetic inductance fraction, which for the same absolute dimensions leaves less room for improvement.
\section{Conclusions}
We performed a systematic study of CPW resonators with a central line width varying between 300 nm and 3.0 $\mum$. It is shown that the responsivity of these resonators increases with reducing width as  $d\theta/dN_{qp} \propto S^{-1.29 \pm 0.04}$. The minimum noise level of the resonators increases due to frequency noise caused by TLS $S_f/f_{res}^2 \propto S^{-1.28\pm0.21}$ and the reduction in maximum read-out signal power that the resonators can handle: $P_{int} \propto S^{2.16 \pm 0.20}$. No deviation from these trends is found for the narrowest submicron resonators investigated in this research. This encourages a study with Al resonators, in which theory does predict an appreciable improvement of the resonator sensitivity for reduced CPW widths.

\begin{acknowledgements}
We would like to thank S.J.C. Yates for his data analysis expertise and N. Vercruyssen, Y.J.Y. Lankwarden and E.F.C. Driessen for their suggestions concerning fabrication. A.~Endo is financially supported by NWO (Veni grant 639.041.023) and JSPS Fellowship for Research Abroad. T.M.~Klapwijk likes to thank the W.M.Keck Institute for Space Sciences  for partial support for his stay at California Institute of Technology, while this manuscript was being written.
\end{acknowledgements}

%\pagebreak

\end{document}